\documentstyle[12pt,epsfig]{article}

\begin{document}
\topmargin -1.4cm
\oddsidemargin -0.8cm
\evensidemargin -0.8cm 

\title{Anti-screening  ferromagnetic superconductivity}

\vspace{1.5cm}

\author{P. Olesen\\
{\it  The Niels Bohr Institute, University of Copenhagen,}\\
{\it Blegdamsvej 17, Copenhagen \O, Denmark }}

\maketitle
\begin{abstract}
We consider the Ginzburg-Landau mean field theory of ferromagnetic
superconductors with a complex vector order parameter $\psi_i, ~i=1,2,3$. 
Below the critical temperature
superconductivity is generated by means of a tachyonic term
in the free energy
-$\alpha |\psi_i|^2 ~~(\alpha >0)$ stabilized by $\psi^4-$terms. However, 
we show that when $\alpha$ becomes negative 
superconductivity can also be generated spontaneously by the 
magnetic field through the Zeeman-like coupling between this field and
the spin due to the order parameter, in analogy with what happens in
the condensation of $W$ mesons. We show that this mechanism leads to
a vortex lattice with anti-screening currents, enhancing the
magnetic field instead of counteracting it. This lattice of vortices is a
collective phenomenon where the individual elements do not exist in isolated 
form. If the parameter $\alpha$ vanishes it is possible to obtain an
explicit form for the vortex lattice in terms of Weierstrass' p-function.
 
\end{abstract}


\thispagestyle{empty}
\newpage

\vskip0.5cm

\section{Introduction}

During the last years ferromagnetic superconductors have been found \cite{1}-
\cite{5} where spin triplet state Cooper pairs exist. In a mean field
Ginzburg-Landau (GL) description the order parameter is therefore a 
complex vector field $\psi_i, 
i=1,2,3$. The Gibbs free energy density \cite{6}-\cite{8} consists of a number 
of terms, 
\begin{equation}
{\cal F}_{\rm grad}=K_1|D_i\psi_j|^2+K_2[(D_i\psi_j)^*(D_j\psi_i)+
(D_i\psi_i)^*(D_j\psi_j)]+K_3|D_i\psi_i|^2,~~~D_i=\partial_i-i2eA_i,
\label{1}
\end{equation}
where $h/2\pi=c=1$,and $2e$ is the charge of the Cooper pair,
\begin{equation}
{\cal F}_{\rm mixed}=i J 4\pi M_i~\epsilon_{ijk}~\psi_j^*~\psi_k+\gamma 
M_i^2~|\psi_j|^2,
\label{2}
\end{equation}
and
\begin{equation}
{\cal F}_{\rm pot}=-\alpha|\psi_i|^2+\beta (\psi_i^*\psi_i)^2+\beta_2
|\psi_i^2|^2.
\label{3}
\end{equation}
Here the potentials $A_i$ are associated with the field  $B_i=H^{ext}_i+
4\pi M_i$, where $H^{ext}_i$ is a possible external magnetic field
and $M_i$ is the magnetization density. The total Gibbs free 
energy density
is then  a sum of the three expressions above plus a pure magnetic free 
energy density for the ferromagnet. The various constants are parameters 
depending on the 
material  as well as on the temperature and the pressure.  The
first term in Eq. (\ref{2}) is the Zeeman-like coupling between the magnetic 
field
and the spin of the Cooper pairs.

In the superconducting state $\alpha$ is usually taken to be positive.
The central point in our paper is that even if $\alpha \leq 0$ there is
a vortex solution and the material is still superconducting in the sense
that the type II London equation is satisfied. However, the currents
are anti-screening, i.e. they enhance the magnetic field instead of
counteracting it. Ultimately this vortex lattice becomes unstable
and disappears because of thermal fluctuations resulting in a 
sufficiently large Lindemann ratio.

The ferromagnetic superconductor with the GL free energy density
(\ref{1})-(\ref{3}) is somewhat similar to
the condensation of $W$ mesons which occurs for large magnetic fields
\cite{r1}. Since the $W$ is represented by a vector field $W_i$ there 
are similarities with the order parameter $\psi_i$, as will be discussed in the
following in more details. However, there are also important differences, in
particular concerning the term $-\alpha |\psi_i|^2$.

In Section 2 we discuss the condensation of $W'$s, and in Section 3 we list 
some of the unusual properties of the resulting $W$-vortex lattice.
In Section 4 we
find a similar vortex lattice solution of the GL-eqs. (\ref{1})-(\ref{3}) for 
$|\psi_i|$. This solution satisfies the London type equation (see Eq. (\ref
{london}) in Section 4)
\begin{equation}
{\rm supercurrent}_k={\rm const.}~A_k|\psi_i|^2.
\end{equation} 
The sign is opposite to the sign in the usual type II superconductor, due
to the fact that we have anti-screening. 
At or very
near the critical temperature this solution can be expressed explicitly in 
terms of Weierstrass elliptic p-function $\wp$, as shown in Section 5. In 
Section 6 we discuss the results obtained.

\section{Magnetically induced $W$ condensation}

We shall now review the most important features of the condensation of $W$ 
mesons in a large magnetic field, first discussed several years ago \cite{r1}.
Except for a few remarks this exposition does not contain new material.
At the end of this section it will then become obvious that there 
are some similarities with the free 
energy, Eqs. (\ref{1})-(\ref{3}). However, as we shall see, there are also 
important differences. 

A phenomenon similar to $W$ condensation has been found to occur for
$\rho$ mesons \cite{rho2}, \cite{rho3}. 

We consider a simple SO(3) model\cite{r1} with an isovector field 
$A_i^a$. Introducing the complex field
\begin{equation}
W_i=\frac{1}{\sqrt{2}}~(A_i^1+iA_i^2),
\end{equation}
we have the static energy \cite{r1}
\begin{equation}
{\cal E}=\frac{1}{4}~F_{ij}^2+m^2|W_i|^2,
\end{equation}
or
\begin{equation}
{\cal E}=\frac{1}{4}~f_{ij}^2+\frac{1}{2}|D_iW_j-D_jW_i|^2+m^2 |W_i|^2
+ief_{ij}~W_i^*W_j+\frac{1}{2}e^2[(W_i^*)^2W_j^2-(W_i^*W_i)^2)],
\label{4}
\end{equation}
where $f_{ij}=\partial_iA_j-\partial_jA_i,~ A_i=A_i^3$ and (in this Section) 
$D=\partial-ieA$. Since we consider a time
independent situation we can take $W_0=0$ and $A_0=0$. We see that there is 
some similarity with the free energies Eqs. (\ref{1})-(\ref{3}). However, 
the mass 
term in (\ref{4}) is positive, in contrast to the $\alpha$ term in (\ref{3}). 
The term $ief_{ij}~W_i^*W_j$ is due to the magnetic moment of the $W$ and is
similar to the Zeeman-like term in Eq. (\ref{2}).

Let us consider a magnetic field $B_3$ in the third direction. It was then 
shown many years ago \cite{nknpo} that there is an instability (growing in 
time, as can be seen by adding a term $-|\partial W_i/\partial t|^2$) for 
large fields $B_3\geq m^2/e$ in the linearized theory, where $W^4$ terms are 
ignored. This instability appears in $W_1$ and $W_2$ such that
\begin{equation}
W_2(x_1,x_2)=iW_1(x_1,x_2)\equiv iW(x_1,x_2),~W_3=0.
\label{unstablew}
\end{equation}
Further, these unstable $W$ fields are solutions of the equation
\begin{equation}
D_iW_i=0,~~~{\rm or}~~~(D_1+iD_2)~W=0.
\label{sub}
\end{equation}
For a constant magnetic field the solution of this equation is 
$W=\exp[-\frac{1}{2}eB_3x_1^2]  ~F(x_1+ix_2)$, where $F$ is an analytic 
function. This solution also corresponds to the eigenfunction for the
imaginary eigenvalue from the time dependent linearized equation of motion 
from (\ref{4}).  From Eq. (\ref{4}) and Eq. (\ref{sub}) we then get the 
equation of motion
\begin{equation}
\left[-(\partial -ie A)^2+m^2-2ef_{12}\right]~W=-2e^2|W|^2W.
\label{eqs}
\end{equation}
From the condition (\ref{sub}) we further have
\begin{equation}
(D_1-iD_2)(D_1+iD_2)W=0,~~~{\rm or}~~~(D_1^2+D_2^2)W=-i[D_1,D_2]W=-ef_{12}W.   
\end{equation}
From the equation of motion (\ref{eqs}) we then get
\begin{equation}
f_{12}=\frac{m^2}{e}+2e|W|^2.
\label{antis}
\end{equation}
We note the signs here. For an Abelian field the corresponding prototype
equation would be
\begin{equation}
f_{12}^{Abelian}=\phi_0^2-|\phi|^2,
\label{s}
\end{equation}
where $\phi$ is a complex scalar field.

Eq. (\ref{sub}) can be used to solve the potentials in terms of $W=
|W|e^{i\chi}$,
\begin{equation}
eA_i=\epsilon_{ij}~\partial_j\ln |W|+\partial_i\chi
\label{a}
\end{equation}
From this and Eq. (\ref{antis}) we get the equation for $|W|$,
\begin{equation}
-(\partial_1^2+\partial_2^2)\ln|W|=m^2+2e^2|W|^2-\epsilon_{ij}\partial_i
\partial_j \chi.
\label{Weq}
\end{equation}
The current is given by \cite{r1}
\begin{equation}
\partial_i f_{ik}=-j_k=-4e^2A_k|W|^2.
\label{current}
\end{equation}
The sign is opposite to the sign one would have in the Abelian case, where
\begin{equation}
\partial_i f_{ik}^{Abelian}=+4e^2A_k|\psi|^2.
\end{equation}
Eq. (\ref{Weq}) allows periodic solutions where the flux is quantized 
\cite{r1},
\begin{equation}
{\rm Flux}=\int f_{12}d^2 x=\int A_idx_i=\frac{2\pi}{e}.
\label{flux}
\end{equation}
The integration is over one periodicity cell. In each cell $W$ has a zero,
and the delta function from the double derivative of ln$|W|$ in the zero is
exactly cancelled by the delta function from the term $\epsilon_{ij}
\partial_i\partial_j \chi$ in Eq. (\ref{Weq}). 

The result is thus that we have a vortex lattice. It is, of course,
non-trivial to demonstrate that there actually is a periodic solution of
Eq. (\ref{Weq}). This has been done numerically \cite{num} and 
mathematically \cite{math}-\cite{math4}, for the simple model above  as well 
as for the more complicated electroweak theory \cite{ew},\cite{ew2} where the 
mass $m$ is generated by the Higgs field. We also mention that recently the 
propagation of fermions in the vortex lattice has been studied \cite{oxford}.

\section{Properties of the $W$ vortex lattice}

We shall now list a number of properties of the $W$ condensate, remembering
that for physical reasons the mass $m$ is positive:

\vskip0.5cm

(A) The non-Abelian vortex lattice is a superconductor with the unusual
property that the current is antiscreening. The magnetic field is enhanced
by this kind of superconductivity, in contrast to conventional superconductors
with the Meissner effect. In the  topological zero of the vortex the order
parameter vanishes. Thus from Eq. (\ref{antis}) it follows that the magnetic 
field has a minimum in this point. However, in the Abelian case from Eq.
(\ref{s}) we see that instead the field has a maximum at the zero of $|\phi|$.

(B) There is no Meissner phase with a homogeneous order parameter $|W|=$const.
This follows from Eq. (\ref{Weq}); A solution $|W|=$const., $\chi=$const.
does not satisfy this 
equation, since the left hand side must vanish whereas the right hand
side is clearly non-vanishing. In a conventional GL superconductor the
right hand side of Eq. (\ref{Weq}) would be replaced by an expression the 
prototype of which is
\begin{equation}
\phi_0^2-|\phi|^2+\epsilon_{ij}\partial_i\partial_j \chi,
\end{equation}
which can vanish for $|\phi|=$const.$=\phi_0$ and the phase $\chi$ fixed.

(C) There is no Meissner effect. Due to the antiscreening the magnetic field 
is always enhanced, not counteracted. Thus, Lenz' law is not valid.

(D) In the $W$ case superconductivity is spontaneously induced by the 
magnetism. Without
a magnetic field above the threshold $m^2/2$ there is no supercondductivity.

(E) The vortex lattice is a collective phenomenon. The vortices do not
exist as individual isolated objects. This is in contrast to the Abelian case
where a single vortex exists with boundary condition $|\phi|\rightarrow 
\phi_0$ at infinite distances, where the magnetic field approaches zero. 

(F) The threshold for superconductivity is very high for the $W$ meson, and can
only occur in cosmologocal settings or, for a short time, in the LHC collider. 
However,
it may be that the vector field is an order parameter for some material, in
which case $m^2$ is just some (positive) parameter. Positivity of $m^2$ is
physically important, of course, but it is also important in order to have 
dominance of the unstable mode, because if $m^2<0$
the mass is tachyonic and causes instability in the linear approximation,
which competes with the instability from the spin coupling. The latter
would then not be dominant, and the analysis leading to the importance of the
unstable mode (\ref{unstablew}) is then not be valid.

\section{An  antiscreening vortex lattice in\\ ferromagnetic 
superconductors}

We shall now turn to the GL mean field theory for the ferromagnetic
superconductor introduced in Eqs. (\ref{1})-(\ref{3}). In this case vortex 
solutions have been discussed in the literature, see refs. \cite{6},\cite{v1}-
\cite{v3}. The term $-\alpha |\psi_i|^2$ in Eq. (\ref{2}) for $\alpha$ positive
is quite similar to the tachyonic term in conventional GL mean
field theory for scalar superconductors. In the linear approximation this term
causes an instability, which is stabilized by the $\psi^4$-terms. Now
our point is that even when this tachyonic behavior disappears because
$\alpha$ becomes negative above some critical temperature, there is still,
in analogy with the $W$ case discussed in Section 2,  an
instability due to the Zeeman like coupling of the magnetic
field with the spin of the vector order parameter in Eq. (\ref{2}). Again,
this will be stabilized by the $\psi^4$ terms.

In this section we shall therefore discuss
what happens when the traditional kind of superconductivity no longer occurs 
because $\alpha <0$. In analogy with the $W$ case a different kind 
of superconductivity is then
created spontaneously from the magnetic field. In the following 
we consider the situation
where there is no external field, so $B_i=4\pi M_i$.

Usually one takes 
\begin{equation}
\alpha=a(T_c-T),~~a>0,
\end{equation}
at least in the neighbourhood of the critical temperature $T_c$. This
implies that conventionally the material is superconducting below $T_c$,
since $\alpha$ is positive in the superconducting state according to
conventional GL theory \cite{6}. However, in the following we 
take $\alpha$ negative
and we shall see that there is still an {\it anti-screening}
superconducting vortex lattice, due to the unstable mode \cite{nknpo} 
discussed before. To see this we proceed like in the $W$ case discussed in
Section 2. In analogy with (\ref{unstablew}) the unstable mode now 
corresponds to
\begin{equation}
\psi_2(x_1,x_2)=i\psi_1(x_1,x_2)\equiv i\psi(x_1,x_2),~~\psi_3=0,
\label{unstable}
\end{equation}
and we impose the condition
\begin{equation}
D_i\psi_i=0,~~{\rm or}~~~(D_1+iD_2)\psi=0.
\label{D}
\end{equation}
Then, as before, we get with $D=\partial-2ieA$
\begin{equation}
(D_1-iD_2)(D_1+iD_2)\psi=0,~~{\rm or}~~(D_1^2+D_2^2)\psi=-2ef_{12}\psi.
\label{100}
\end{equation}

In the following we take $\gamma=0$ in order to neglect the term
quadratic in $M_i$ in $\cal F_{\rm mixed}$ in Eq. (\ref{2}). This
means that in the linear approximation an instability arises from the
Zeeman-like term in $\cal F_{\rm mixed}$.
We can now  minimize the free energy density in Eqs. (\ref{1})-(\ref{3}),
\begin{equation}
-K_1D_i^2\psi_j+2ieK_2 f_{ij}\psi_i+iJB_i\epsilon_{ijk}\psi_k-\alpha\psi_j+
2\beta (\psi_i^*\psi_i)\psi_j+2\beta_2\psi_i^2\psi_j^*=0.
\end{equation}
Using the unstable mode (\ref{unstable}) this reduces to
\begin{equation}
-K_1(D_1^2+D_2^2)\psi-(J-2eK_2)f_{12}\psi=\alpha\psi-4\beta|\psi|^2\psi.
\end{equation}
Using Eq. (\ref{100}) we finally obtain the expression
\begin{equation}
(J-2eK_1-2eK_2) B_3=-\alpha+4\beta|\psi|^2.
\label{antisy}
\end{equation}
It should be remembered that we are above the critical temperature so 
$\alpha$ is negative. Also, we assume that
\begin{equation}
J>2eK_1+2eK_2.
\end{equation}
If this is not the case the above analysis will have to be redone with a
different unstable mode.

The free energy density should also have a magnetic part. Since we do not have
an external field we take
\begin{equation}
{\cal F_{\rm magnetic}}=\frac{1}{4}~f_{ij}^2+...,~~~f_{ij}=\partial_iA_j-
\partial_jA_i=\epsilon_{ijk}B_k.
\end{equation}
Here the dots stand for possible higher order terms in $f_{ij}$. We assume that
the magnetic field is so small that these terms can be ignored relative to
the quadratic term.
From Eq. ({\ref{antisy}}) we obtain the current by use of 
$\partial_if_{ik}=-j_k$,
\begin{equation}
j_k=\frac{4\beta}{J-2eK_2-2eK_1}\epsilon_{ki}\partial_i|\psi|^2=
\frac{8e\beta}{J-2eK_2-2eK_1}~A_k|\psi|^2.
\label{london}
\end{equation}
Again we see that the sign is the opposite of what it is in conventional
scalar superconductors. Here we used that the condition (\ref{D}) can be 
solved for the potentials to give
\begin{equation}
2eA_i=\epsilon_{ij}\partial_j\ln |\psi| +\partial_i\chi
\end{equation}
where the phase of $\psi$ is denoted $\chi$. For the purpose of computing the
current $j_i$ this phase has been absorbed as a gauge which is allowed away
from the topological zeros in $\psi$, where the current vanishes. The flux
is computed like in Eq. (\ref{flux}),
\begin{equation}
{\rm Flux}=\int d^2x B_3=\int_{\rm cell}A_idx_i=\frac{\pi}{e}.
\end{equation}
It should be remembered that the charge of a Cooper pair is $2e$.

Finally, we have the equation of motion fot $|\psi|$,
\begin{equation}
-(\partial_1^2+\partial_2^2)\ln|\psi|=\frac{1}{2e(J-2eK_2-2eK_1)}~(-\alpha+
4\beta |\psi|^2)-\epsilon_{ik}\partial_i\partial_k\chi.
\label{eofm}
\end{equation}
The proofs of existence of periodic solutions found in \cite{math}-
\cite{math4} are still
valid, since we take $\alpha$ negative, corresponding to a temperature
higher than $T_c$. Thus $-\alpha$ is like the square of a mass.

The conclusion is thus that even if superconductivity naively disappears
for $\alpha<0$ it is spontaneously generated by the magnetic field
even above the critical temperature. The resulting vortex lattice looks
superficially like an Abrikosov lattice, but by closer inspection we
see that the currents ant-screen the magnetic field in contrast to what happens
in the Abrikosov lattice where there is screening.

\section{An explicit solution at the critical temperature}

We end by pointing out that at the critical temperature where $\alpha=0$ there
exist an explicit solution of the eqution of motion (\ref{eofm}). For other
values of the parameter  $\alpha$ it is necessary to solve this equation
numerically. 

For $\alpha=0$ Eq. (\ref{eofm}) reduces to the Liouville equation
\begin{equation}
-(\partial_1^2+\partial_2^2)\ln|\psi|=\frac{2\beta~ |\psi|^2}
{e(J-2eK_2-2eK_1)}.
\end{equation}
This equation has been solved with periodic boundary conditions \cite{solve1}-
\cite{solve3},
\begin{equation}
|\psi|=\sqrt{\frac{2e(J-2eK_2-eK_1)}{\beta (e_2-e_3)(e_3-e_1)}}~
\frac{|\wp'(z)| }{1+|\wp (z)-e_3|^2/((e_3-e_1)(e_2-e_3))}.
\label{weier}
\end{equation}
Here $\wp$ is Weierstrass' doubly periodic function, $z=x_1+ix_2$, and the 
$e'$s are the roots of
\begin{equation} 
4t^3-g_2t-g_3=0,
\end{equation}
where the $g'$s are defined in the standard literature on Weierstrass' 
p-function. We assume thet these roots are real (requires $g_2^3-27g_3^2>0$) 
and $e_2>e_3>e_1$. Other forms 
of the solution can be found in \cite{solve2} and \cite{solve3}. In 
Eq. (\ref{weier}), if the function $\wp$ is doubly periodic with periods
$2a,2ib$, then $|\psi|$ is periodic with periods $a,ib$. We note that if
some material exists with $\alpha=0$ the vortex lattice is aleays described
by Eq. (\ref{weier}).

\section{Discussion}

We have shown that although ``conventional'' superconductivity 
disappears above the critical temperature $T_c$ nevertheless
it is possible for a magnetic field to generate superconductivity 
spontaneously. The new superconductivity anti-screens the magnetic field,which 
is therefore enhanced in certain regions. For this effect to work the
following threshold condition must be satisfied,
\begin{equation}
B_3\geq \frac{-\alpha}{J-2eK_1-2eK_2}.
\end{equation}
At or near the critical temperature this threshold vanishes or is very
small. It is only when the temperatutre increases that the threshold
becomes essential. 

The properties (A)-(E) mentioned in Section 3 are also valid for the
ferromagnetic superconductor discussed in Sections 4 and 5, so we shall
not repeat them here. 

It is interesting that the effects valid for the ferromagnetic
superconductor are present in the
QCD vacuum, where the magnetic field is generated by quantum corrections
\cite{qcd1} and subsequently introduces a condensate \cite{nknpo}, \cite{qcd2}-
\cite{qcd4}. 

In our discussion of the superconductivity above the critical temperature
we have assumed that $\gamma=0$. If, however, the second term in 
$\cal F_{\rm mixed}$ in Eq. (\ref{2}) is present, the situation changes 
somewhat, since for large enough magnetic fields 
this term will counteract the Zeeman-like term if $\gamma>0$.
Let us for simplicity take the temperature to be close to $T_c$ so
$\alpha$ can be ignored. The condition for the unstable mode to be operative 
is then
\begin{equation}
\gamma B_3< J-2eK_1-2eK_2.
\label{u}
\end{equation}
Thus, in the neighborhood of the critical temperature the magnetic field is 
limited from above. 

We have seen that superconductivity with the generation of a spontaneous
vortex lattice can exist beyond the critical temperature corresponding to 
$\alpha=0$. It may, however,
be that the thermal fluctuations can result in a suufficiently
large Lindemann ratio,
thereby producing a ``spaghetti'' liquid vortex state somewhat similar  to
what was discussed for the QCD vacuum \cite{qcd2}. The property of 
anti-screening would, however, still be maintained. 

We end with the remark that (ferro-)magnetic superconductivity can exist
in a GL approach even without the $-\alpha |\psi_i|^2$ term. If $\alpha=0$ 
superconductivity is induced entirely by the Zeeman like term in Eq. 
(\ref{2}), and we have the explicit solution in terms of Weierstrass' 
p-function given by Eq. (\ref{weier}).
It would be interesting to see experimentally if
such a special vortex lattice exists for some material.

\end{document}